\documentstyle[12pt]{article}
\begin{document}

\def\lsim{\mathrel{\rlap{\lower3pt\hbox{\hskip0pt$\sim$}}
    \raise1pt\hbox{$<$}}}         
\def\gsim{\mathrel{\rlap{\lower4pt\hbox{\hskip1pt$\sim$}}
    \raise1pt\hbox{$>$}}}         

\begin{titlepage}
\renewcommand{\thefootnote}{\fnsymbol{footnote}}

\begin{flushright}
TPI-MINN-95/5-T\\
UMN-TH-1332-95\\
hep-ph/9503404\\
\end{flushright}
\vspace{.6cm}
\begin{center} \LARGE
{\bf Do Higher Order Perturbative Corrections Upset $|V_{cb}|$ and
$|V_{ub}|$ Determined from Semileptonic Widths?}
\end{center}
\vspace*{.4cm}
\begin{center}
{\Large
N.G. Uraltsev}
\vspace*{.6cm}\\
{\normalsize
{\it Theoretical Physics Institute, Univ. of Minnesota,
Minneapolis, MN 55455}\\
and\\
{\it Petersburg Nuclear Physics Institute,
Gatchina, St.Petersburg 188350, Russia}\footnote{Permanent
address}
\vspace*{1.2cm}}\\
{\Large{\bf Abstract}}
\end{center}

\vspace*{.5cm}

It is shown that large perturbative corrections found previously for
semileptonic beauty and charm decays are associated with using
inappropriate pole masses. The latter, in the perturbative expansion, suffer
from the $1/m_Q$ infrared renormalon which is absent in the widths, which
leads to similar large corrections in $m_Q$. Pole masses are neither measured
directly in experiment. If the widths are related to parameters determined in
experiment, the overall impact of the calculated second order corrections
becomes strongly suppressed and leads to less than $1\%$ change in $|V_{cb}|$
and $|V_{ub}|$. Even in charm decays the perturbative corrections appear to be
very moderate in the consistent OPE-compliant treatment. The updated estimate
of $|V_{cb}|$ is given, based on recent accurate determination of $m_b$ and
$\alpha_s(1\,\rm GeV)$. The theoretical accuracy of determination of
$|V_{ub}|$ from $\Gamma_{\rm sl}(b\rightarrow u)$ appears to be
good as well.

\end{titlepage}
\addtocounter{footnote}{-1}

\newpage

Inclusive semileptonic decays of $B$ mesons provide at present the most
accurate determination of the CKM mixing element $|V_{cb}|$ and, if one could
measure the inclusive width $\Gamma(B\rightarrow X_u\ell \nu)$,
the value of $|V_{ub}|$ as well. In order to reach the ultimate possible
theoretical precision it is practically necessary to account for purely
perturbative corrections as accurately as possible. The first order
corrections are
taken directly from the QED calculations for $\mu$ decay known since 50s
\cite{muon}.
Recently the part of the second order corrections has been computed, which is
associated with the running of the strong coupling (we will refer to such
approximation below as to the BLM approach). Let us introduce the following
general notations:
\begin{equation}
\frac{\Gamma_{\rm sl}(m_q/m_b)}{\Gamma^0_{\rm sl}(m_q/m_b)}
=\left(1+a_1\frac{\alpha_s}{\pi}
+a_2\left(\frac{\alpha_s}{\pi}\right)^2 \;+...\right)\;\;\;.
\label{2}
\end{equation}
Then the result for $\Gamma_{\rm sl}(b\rightarrow q\,\ell \nu)$ \cite{wise}
reads (for $\alpha_s(m_b)$ in the $V$ scheme \cite{blm} and $n_f=3$)
$$
a_1=-2.41\;\;,\;\;\;a_2=-19.7\;\;\;\;{\rm at}\;\; m_q/m_b=0
$$
\begin{equation}
a_1=-1.67\;\;,\;\;\;a_2=-8.8\;\;\;\;{\rm at}\;\; m_q/m_b=0.3
\label{1}
\end{equation}
where the first line corresponds to the $b\rightarrow u$ decays and the second
one is relevant for the $b\rightarrow c$ transitions.
In Eq.~(\ref{1}) only the BLM part of $a_2$ is shown. In the absence of the
complete calculations we will discuss only this part to the rest of this paper
without explicit reminding this reservation. It is worth noting
that the values of
$a_2$ quoted in Eq.~(\ref{1}) correspond to the $V$ scheme which has the direct
physical meaning in the BLM calculations \footnote{$\overline{MS}$ scheme is
unnatural here; in this scheme one would have $a_2^{(\overline{MS})} =
a_2^{(V)}+5b/12\,a_1$ where $b=11-2/3n_f\simeq 9$ is the first coefficient in
the strong coupling $\beta$-function.}.
For this reason we will consistently use below the $V$ scheme.

The above values of $a_2$ are surprisingly large at first sight and could be
thought to change significantly the theoretical estimates for the semileptonic
widths, in particular in the $b\rightarrow u$ decay, once $(\alpha_s/\pi)^2$
terms are accounted for. It appears, however, that the actual impact
of these terms is essentially smaller, because similar large
$(\alpha_s/\pi)^2$ corrections affect the values of the heavy quark masses.
This ``conspiracy'' is the reflection of the dominance of the leading $1/m_Q$
infrared renormalon which, in reality, is absent in the widths once they are
expressed in terms of observable quantities \cite{pole}.
The numerical situation with perturbative corrections in semileptonic beauty
decays is, therefore, rather similar to the case of $t\rightarrow b+W$ width
\cite{Volt}.

\section{Heavy Quark Masses and Perturbative Corrections to Widths}

Heavy quark masses are not directly observed and must be determined indirectly
from independent measurements. At present the most accurate determination of
$m_b$ follows from the sum rules for $\bar b b $ threshold production in $\rm
e^+e^-$ annihilation \cite{Volmb}; the value of $m_b$ there undergoes
perturbative corrections as well and this must be accounted for properly.

The typical momentum scale in the analysis of moments of the spectral density
in Ref.~\cite{Volmb} is about $2\,\rm GeV$, and, therefore, these sum rules are
sensitive to the mass normalized at the scale $\mu\gsim  1\,\rm GeV$. It does
not prevent expressing the final result in terms of the ``one loop pole mass''
\begin{equation}
m_{\rm pole}^{(1)} \simeq m(\mu) +\frac{-dm(\mu)}{d\mu}\,\mu
\simeq m(\mu) + c_m\frac{\alpha_s(\mu)}{\pi} \,\mu
\label{5}
\end{equation}
as long as the one loop corrections are concerned.
However, accounting for the $\alpha_s^2$ terms shifts the corresponding
value of the
pole mass to the value of the ``two loop pole mass'' which is given in the BLM
approximation by
\begin{equation}
m_{\rm pole}^{(2)}\simeq m(\mu) + c_m\frac{\alpha_s(\mu)}{\pi} \,\mu+
c_m\frac{b}{2} \left(\frac{\alpha_s(\mu)}{\pi}\right)^2 \,\mu
= m_{\rm pole}^{(1)}+c_m\frac{b}{2} \left(\frac{\alpha_s(\mu)}{\pi}\right)^2
\,\mu\;\;\;.
\label{6}
\end{equation}
Numerically this increase is as large as $150\,\rm MeV$ and this
strongly affects the width as well. (Eqs.~(\ref{5}) and (\ref{6})
assume that $\mu\ll m_Q$.)

It is more practical, both theoretically \cite{pole} and phenomenologically,
to reexpress even the perturbative relation for
the width in terms of the running masses
normalized at the scale $\mu$. To the second order in $\alpha_s$ it would
require the explicit knowledge of the complete two loop scale
dependence of the heavy quark mass, which is not known yet \footnote{The two
loop running of $m_Q$ is known in the dimensional regularization, which,
however, is irrelevant for $\mu\ll m_Q$. $m_Q(\mu)$ in naive dimensional
regularization has nothing common to the properly infrared defined mass needed
for performing OPE \cite{pole}.}. Being interested only in the BLM-type
contribution to
$a_2$, on the other hand, we need to account only for the terms
$\sim b (\frac{\alpha_s(\mu)}{\pi})^2$ which are easily obtained from the first
order calculation. This relation becomes especially simple in the heavy quark
expansion \cite{pole} which holds when $\mu \ll m_Q\,$:
\begin{equation}
m_{\rm pole}\simeq m(\mu) + c_m\frac{\alpha_s(\lambda)}{\pi} \,\mu+
c_m\frac{b}{2} \left(\frac{\alpha_s(\lambda)}{\pi}\right)^2
\left(\log{\frac{\lambda}{\mu}}+1\right)\,\mu\; +\;...
\label{7}
\end{equation}
where $\lambda$ is an arbitrary renormalization scale for $\alpha_s$ (in the
$V$ scheme). Then one finds for the $b\rightarrow u$ case
$$
\tilde{a}_1(\mu)=-\frac{2}{3}\left(\pi^2-\frac{25}{4}\right)
+5\frac{\mu}{m_b}c_m
$$
\begin{equation}
\tilde{a}_2(\mu)= a_2+ 5\frac{\mu}{m_b}c_m \frac{b}{2}
\left(\log{\frac{\lambda}{\mu}}+1\right)\;\;;
\label{8}
\end{equation}
$\tilde{a}(\mu)$ refer to the perturbative coefficients utilizing running
masses
(for simplicity we do not change the argument of $\alpha_s$ here).
Adopting for the purpose of illustration $\mu=1\,\rm GeV$ and
$c_m=\frac{4}{3}$ (see \cite{pole}), we get for $\lambda=m_b$
\begin{equation}
\tilde{a}_1(\mu)\simeq -1.03\;\;,\;\;\;
\tilde{a}_2(\mu)\simeq  -3.7\;\;\;.
\label{9}
\end{equation}
The perturbative corrections appear to be of ``normal'' magnitude.

A similar consideration can be applied to the $b\rightarrow c$ decays, with
only minor technical complications. Here one is relatively close to the SV
limit where the total width would depend only on $m_b-m_c$ rather than on the
absolute values of masses. Then the impact of using the pole masses instead of
the
running ones with $\mu\ll m_c\,,\:m_b$ is suppressed, and one gets essentially
smaller values of $a_2$ from the very beginning. In particular, if one uses the
strong coupling normalized at the scale $\sqrt{m_cm_b}$ which is appropriate
for the SV kinematics \cite{NU,look} then the value of $a_2$ in the $V$ scheme
is
$$
a_2\simeq -4.3\;\;.
$$
This coefficient, still somewhat enhanced, {\em per se} would lead to the
increase in the value of $|V_{cb}|$ by $2$ percentage points. On the other
hand, according to Ref.~\cite{vcb}, just
the increase in the input value of $m_b$ by
$150\,\rm MeV$ leads to the decrease of $|V_{cb}|$ by $2.5$ percentage points.
Therefore when one consistently proceeds from the first order approximation to
the second one, the two effects tend to strongly offset each other.
We will show below that the quoted value of $a_2$ is indeed largely due to
the residual dependence of the
tree level width on the absolute values of masses for actual $m_c$ and $m_b$
and, therefore, in reality the impact of the second order corrections is
further suppressed.

To proceed from the pole masses to $m_c(\mu)$ and $m_b(\mu)$ one can follow the
same way as for $b\rightarrow u$, however in this case the leading in
$\mu/m_c$ approximation used in Eqs.~(\ref{5})--(\ref{7}) {\em a priori}
could be not well justified. The corrections can be readily taken into account
and, as a matter of fact, appear to be small: the deviation from the linear
$\mu$ dependence is less than $7\%$ even for $\mu/m_c=1$. Let us briefly
discuss the general case of arbitrary $\mu/m_Q$.

In the framework of the BLM approach the exact $\mu$ dependence is given by
the value of the one loop diagram evaluated with the running $\alpha_s(k^2)$:
\begin{equation}
m_Q(\mu)-m_Q(\mu')=\int_{\mu^2}^{\mu'^2}\, \frac{\alpha_s(k^2)}{\pi}\,
F_m(k^2) \:dk^2
\label{12}
\end{equation}
where $ F_m(k^2)$ is obtained by integrating the expression for the one loop
Feynman graph over the directions of the gluon momentum in the 4 dimensional
Euclidean space. In fact, this quantity is conveniently expressed in terms of
the one loop correction evaluated with the massive gluon propagator: assigning
the gluon mass $\nu$ one calculates, say, $\delta m^{(1)}_Q(\nu^2)$
substituting $1/k^2 \rightarrow 1/(k^2+\nu^2)$ in the gluon
propagator. Then $k^2F(k^2)$ for any observable
is equal to the discontinuity of the corresponding $\nu^2$-dependent
quantity with respect
to $\nu^2$ at
the negative value of the gluon mass squared. For example, for the heavy quark
mass one has
\begin{equation}
 \frac{\alpha_s}{\pi}\cdot k^2F_m(k^2)= - \frac{1}{2\pi i}
\left(m_Q^{(1)}(\nu^2=-k^2+i\epsilon)-m_Q^{(1)}
(\nu^2=-k^2-i\epsilon)\right)\;\;.
\label{13}
\end{equation}
Eq.~(\ref{13}) thus expresses the generic function $w$
considered recently
in Ref.~\cite{flow} in terms of the gluon mass dependent observable which had
been introduced
in Refs.~\cite{smvol,bbz,bb2}.

Expanding Eq.~(\ref{12}) in $\alpha_s(\lambda)$ one gets
\begin{equation}
m_{\rm pole}=m_Q(\mu) + \frac{\alpha_s(\lambda)}{\pi} \int_0^{\mu^2}
F_m(k^2)\,dk^2\;+ \frac{b}{4}\left( \frac{\alpha_s(\lambda)}{\pi}\right)^2
\int_0^{\mu^2} F_m(k^2)\,\log{\frac{\lambda^2}{k^2}}\,dk^2\;+\;...
\label{14}
\end{equation}
(the pole mass to any finite order is defined as $\,\lim_{\mu\rightarrow
0}m_Q(\mu)\,$).

At $\nu^2\ll m_Q^2$ one has \cite{BU,pole} $m_Q^{(1)}(\nu^2)-m_Q^{(1)}(0)
\simeq -\frac{2}{3}\alpha_s
\,\sqrt{\nu^2}$ and, thus, $F_m(k^2)\simeq 2/(3\sqrt{k^2})$ which yields
Eqs.~(\ref{5}), (\ref{7}) with $c_m=4/3$. The correction factors
to these approximate relations are
easily calculated using the exact one loop expression for $m_Q^{(1)}(\nu^2)\,$:
\begin{equation}
m_Q^{(1)}(\nu^2)=m_Q(0)-\frac{2}{3}\frac{\alpha_s}{\pi}m_Q \left[
\int_0^1\:d\alpha(1+\alpha)\,\log{\left(
1+\frac{\nu^2}{m_Q^2}\frac{1-\alpha}{\alpha^2} \right)}
\right]\;\;.
\label{mass}
\end{equation}
The integral is computed straightforwardly (see, e.g. \cite{bb2}) but
is inessential for us. The discontinuity over $\nu^2$ is obtained
directly from Eq.~(\ref{mass}) and takes the form
\begin{equation}
F_m(k^2)\equiv \frac{1}{2m_Q}\frac{{\cal F}_m(t)}{\sqrt{t}}=
\frac{1}{3m_Q}
\left( \left(1-\frac{t}{2}\right)
\sqrt{1+\frac{4}{t}}+\frac{t}{2}
\right) \cdot\theta(t)\;\;,
\label{fm}
\end{equation}
$$
{\cal F}_m(t)\simeq \frac{4}{3}-\frac{t}{2}+\frac{t^{3/2}}{3}\,-\:...
\;\;\;,\;\;\;\;\;\; t=\frac{k^2}{m_Q^2}\;\;\;.
$$
This coincides with the expressions quoted in Refs.~\cite{bb2,flow,bbb}.
The correction factors corresponding to integration of the exact function
${\cal F}_m$ instead of its approximate value $4/3$ for small $k^2$, differ
from unity by only $0.075$ and $0.05$ for the first and second orders in
$\alpha_s$, respectively, even at $\mu=m_Q$.

Taking all these corrections into account and using
\begin{equation}
\Gamma_{\rm sl}=\frac{m_b^5 |V_{qb}|^2}{192\pi^3}
z_0\left(\frac{m_q^2}{m_b^2}\right)\left(1+a_1\frac{\alpha_s}{\pi}+
a_2\left(\frac{\alpha_s}{\pi}\right)^2\;+\;
... \right)
\label{15}
\end{equation}
with
$$
z_0(x)=1-8x-12x^2\log{x}+8x^3-x^4
$$
one gets
$$
\tilde{a}_1(\mu) =
-1.03\;\;,\;\;\;\tilde{a}_2(\mu) = -3.7\;\;\mbox{ for }\;b\rightarrow u
$$
\begin{equation}
\tilde{a}_1(\mu) = -1.07\;\;,\;\;\;\tilde{a}_2(\mu) =
0.9\;\;\mbox{ for }\;b\rightarrow c
\label{16}
\end{equation}
for $\mu=1\,\rm GeV$, $m_b=4.8\,\rm GeV$ and $m_c/m_b=0.3$ (as previously, I
use the $V$ scheme $\alpha_s$ normalized at $m_b$ for $b\rightarrow u$ and at
$ \sqrt{m_c m_b}$ for $b\rightarrow c$). The values of the
second order coefficients are moderate and $\tilde{a}_1(\mu)$ are
suppressed as compared to the case of the pole masses.

Let us emphasize that the estimates above were for illustration purposes only;
to show the dominant source of the perturbative coefficients I deliberately
limited myself only to the redefinition of masses while not changing the
scale of $\alpha_s$ or the {\em numerical} values of representative quark
masses. These additional modifications are accounted for properly in the
final numerical evaluation; not surprisingly, they do not change the width by
any noticeable amount.

We now dwell on the definition of the renormalization point $\mu$ for masses
used above. The separation of low and high momenta can be accomplished
technically in different ways leading, e.g., to somewhat different values of
the coefficient $c_m$. The most natural choice for nonrelativistic expansion is
the cutoff over the spacelike gluon momenta, $|\vec k|>\mu$, used in
Ref.~\cite{optical} for heavy flavor transitions in the SV kinematics. It
is also most natural for the nonrelativistic $b \bar b$ in the $\Upsilon$
system. This method applied to calculation of the heavy quark mass
yields the value $c_m=4/3$ used above. In general, in the
BLM-type calculations the step-like cutoff over Euclidean four momentum
$\theta(k^2-\mu^2)$ acts differently; however, in the case of the heavy quark
mass $\mu$-dependence the
two approaches are equivalent in the leading in $\mu/m_Q$
approximation~\footnote{A word of caution is worthwhile.
If one uses not a Lorentz invariant cutoff then the
relativistic dispersion law can be modified. For example, the effective heavy
quark mass
which determines the rest frame energy may start to differ
from the one in the dispersion law $E(\vec q)-E(0)=\vec{q}\,^2/2m_Q+...\;$. It
is for this reason the slightly different value of $c_m=16/9$ was obtained in
Refs.~\cite{Volopt,optical} considering the recoil effects in the
second sum rules
in the SV limit. I am grateful to M.~Voloshin for the discussion of this
point. In the concrete version of the BLM approach considered in the
present paper this subtlety is absent.}.

It was found in Ref.~\cite{Volmb}, in accord with the general arguments above,
that the one loop value of the $b$ quark pole mass $m_{\rm pole}^{(1)}$ was
strongly correlated with the obtained value of $\alpha_s$. On the other hand,
the one loop pole mass
normalized at $1.3\,\rm GeV$,
$$
m_{\rm pole}^{(1)}- \frac{4}{3} \frac{\alpha_s}{\pi}\cdot 1.3\,\rm GeV
$$
was essentially uncorrelated with $\alpha_s$ and literally had
significantly smaller
range of variation \cite{Volmb}. Therefore, such a normalization scale can be
considered as physically appropriate one; at least, it eliminates
the strong correlation to the first loop approximation.
The possible
remaining (not enhanced) second order corrections to
the mass normalized at this scale are then the part of the theoretical
uncertainty of determination of the heavy quark mass in the approach of
Ref.~\cite{Volmb}. Clearly, one can, in principle, use arbitrary normalization
point $\mu$ of approximately this scale.
The dependence of $\tilde{a}_1$ and $\tilde{a}_2$ on $\mu$ at $c_m=4/3$ is
illustrated in Table~1. Perturbative corrections appear to be suppressed for
any reasonable choice of $\mu$.

The following remark is in order here. The value of the first order coefficient
is also noticeably suppressed if one uses the running masses $m_Q(\mu)$
above
instead of the pole masses. Accordingly, the value of $\tilde{a}_2$ is less
sensitive to the particular scheme for $\alpha_s$. For example, even in the
$\overline{\rm MS}$ scheme the values of $\tilde{a}_2$ for $\mu=1.3\,\rm GeV$
become $-3.4$ and $-1.5$ for $b\rightarrow u$ and $b\rightarrow c$,
respectively, i.e. small as well. Moreover, it is clear that the value of the
whole perturbative factor through the second order,
$1+\tilde{a}_1\alpha_s/\pi+\tilde{a}_2(\alpha_s/\pi)^2$ does not exhibit any
noticeable dependence on the scheme chosen for $\alpha_s$
(which is always present at some level due to the
truncation of the series).

The above analysis for the $b\rightarrow u$ case shows that practically the
whole
large second order contribution obtained in Ref.~\cite{wise} numerically comes
from the domain of gluon momenta $\sim (0.2\div 0.25) m_b$ and, therefore, is
mainly
associated with the leading $1/m_Q$ renormalon in the pole mass which, however,
is absent in the width in the proper treatment \cite{pole,bbz}. Applying this
observation to the semileptonic charm decays we identify that the large
``nonconvergent'' second order correction pointed out in \cite{wise} came,
in fact, from the gluon momenta
about and {\em below} the infrared pole in the gluon propagator; this clearly
explains the observed numerical behavior. This contribution is, therefore,
phenomenologically irrelevant and is explicitly excluded even from perturbative
corrections in the consistent treatment. In particular, for $\mu\simeq
400\,\rm MeV\,$,
which seems to be the minimal reasonable cutoff, the perturbative corrections
for charm literally become only
$$
1-0.5\frac{\alpha_s}{\pi}-0.4\left( \frac{\alpha_s}{\pi}\right)^2\simeq
1-0.07-0.008\simeq 0.92\;\;;
$$
the second order yields a tiny contribution only. Altogether one has
very moderate
effect \cite{look}. The non-BLM higher order terms are potentially more
important,
but most likely are by far dominated by calculable power nonperturbative
\cite{buv} and unknown ``exponential'' terms \cite{dike}.

We see that in the conventional approach, when all intermediate results are
expressed in terms of the pole mass, the final result for the width is
rather sensitive to the treatment of the infrared region in the Feynman
integrals; already at an accuracy level of $5\div 10\%$ in beauty decays care
must be taken to do it in a self-consistent way. On the other hand,
just the physics at this scale affects the widths very little \cite{buv,pole}
(as long as one
keeps the high scale heavy quark masses fixed rather than the masses of
observable hadrons). Therefore it is advantageous to get rid of this
low energy region and, therefore, of the
pole mass in the extraction of weak mixing parameters from start.
It is very
important from phenomenological point of view that the sum rule analysis of the
$b \bar b $ production \cite{Volmb} allows the direct determination of the high
scale $m_b(\mu)$ with $\mu$ in the interval $ \sim (0.75\div 1.5) \,\rm GeV$.
The theoretical accuracy there is optimal for $\mu \approx 1\,\rm GeV$,
in accord with the general OPE picture, where the impact of low momentum
region is described by the vacuum expectation value of
$G_{\alpha\beta}^2(\mu)$ and
was estimated to be negligible \cite{Volmb} for any
reasonable normalization scale.

This fact is crucial for
reliability of theoretical calculations, because the perturbative treatment in
the region where $\alpha_s$
is not small is rather involved and, as a matter of fact, is sensitive to the
various approximations. For example, going beyond the BLM computations can
change drastically all corrections coming from this domain, both for the pole
mass and for the width expressed through it.
The situation is quite different if the low momentum region is explicitly
excluded. In this case the higher order terms produce small impact. This will
be numerically illustrated in Sect.~2 when the result for $|V_{qb}|$
changes only by a minute amount due to
the second order corrections, when expressed
in terms of $m_Q(1\,\rm GeV)$. Only then one can count on smallness of
higher order
corrections, and on the moderate size of corrections due to the residual
$\alpha_s^2$ terms not captured in the BLM approximation.

We need to emphasize here
two points related to the above discussion. First, in the
Wilson procedure of treatment the strong coupling domain of QCD one is to
cut the perturbative
integrals at the scale $\mu$ to ensure the reasonably small value of
the coupling at $\mu$ and above. On the other hand, it is not necessary {\em
per se} to take $\mu$ as low as possible even for purely perturbative
calculations: the
operator expansion can be applied to the perturbative fields as well,
expressing the contribution of the low momentum region in the perturbative
loop diagrams in terms of the {\em perturbative} matrix elements. Moreover, in
such a way the contribution of this region is most simply evaluated and even
computed (see, for example, papers \cite{pole,optical} where this idea was used
in the heavy quark expansion); following this strategy
one can easily go beyond the first loop, or with
reasonable complication, even
beyond the BLM approximation, which normally is
difficult for complete analytical calculations. For example, the characteristic
scale in $b\rightarrow c$ is set by $m_b-m_c$ or $2m_c$, which is rather large,
and thus safely allows for treatment of the domain up to $1\,\rm GeV$ in this
way.

As an example, we can consider the expansion of the pole mass as the lowest
eigenvalue of the rest frame Hamiltonian \cite{optical} obtained in $1/m_Q$
approach to some order $k$ in perturbation theory:
\begin{equation}
m_Q^{\rm pole}- m_Q(\mu) =
\sum_{n=0}^{\infty}\:\lambda_n(\mu)\frac{1}{m_Q^n(\mu)}=
\overline{\Lambda}(\mu)+\frac{\mu_\pi^2(\mu)}{2m_Q(\mu)}\;+\;...
\label{massexp}
\end{equation}
where the series runs in inverse powers of $m_Q(\mu)$. According to
Eq.~(\ref{massexp}), the perturbative matrix elements $\lambda_n(\mu)$ are
given in the BLM approximation by the corresponding
coefficients of the expansion
of ${\cal F}_m(t)$ (it is defined in Eq.~(\ref{fm})) at small $t$: for
$$
{\cal F}_m(t)=\sum_{n=0}^{\infty}\:c_n t^{\frac{n}{2}}
$$
one has
$$
\lambda_n(\mu)= c_n \frac{\alpha_s(\mu)}{\pi}\:\sum_{l=0}^{k}\:
\int_0^\mu\: d\rho \,\rho^n \left(\frac{b\alpha_s(\mu)}{2\pi}\right)^l\,
\log^l{\frac{\mu}{\rho}}\;=
$$
\begin{equation}
=\;\frac{c_n}{n+1}\frac{\alpha_s(\mu)}{\pi}\,\mu^{n+1}\:\sum_{l=0}^{k}\:
\left(\frac{b\alpha_s(\mu)}{2(n+1)\pi}\right)^l\,l!
\;\;.
\label{pert}
\end{equation}
For $n=0$ we have $\lambda_0(\mu)=\overline{\Lambda}(\mu)$ and, therefore, its
$\mu$-dependence is indeed given by Eq.~(\ref{7}).
Turning to the next term in $1/m_Q$, $n=1$,  we see that ${\cal F}_m(t)$ does
not contain the term
$\sim t^{1/2}$, i.e. $c_1=0$. It means that the matrix element of the kinetic
operator does not receive the perturbative contribution in the BLM
approximation to any order in $(b\alpha_s/\pi)$ when the cutoff is introduced
in
this particular way (this fact was pointed out in the somewhat different
context in Ref.~\cite{bbz}).
Literally the same technique can be applied to other quantities, not only
$m_Q^{\rm pole}$, as well.

The second comment concerns the concrete way of eliminating the low energy
momentum region implicitly implied in our reasoning. Clearly, in
general it does not reduces only to using the running quark masses: even
after expressing the width in terms of $m_Q(\mu)$ integration in Feynman
graphs determining the width runs, strictly speaking, over all momenta
starting $k^2=0$.
Therefore formally an additional
subtraction of the remaining contribution below $\mu^2$ is necessary. It is
easy to see using the above arguments,
however, that in this particular case the situation is rather simple:
these extra terms can be neglected. Indeed, we can use the operator relation
for
the width~\cite{buv,pole}
$$
\Gamma_{\rm sl} \propto \;
m_b^5(\mu)\cdot z\left(\frac{m^2_c(\mu)}{m^2_b(\mu)}\right)
\left[a^{(0)}_{\rm pert}\left((m_c^2/m_b^2);\,\mu\right)
\cdot \left(1-\frac{\mu_\pi^2-\mu_G^2}{2m_b^2}\right)\;- \right.
$$
\begin{equation}
\left.
-\;
a_{\rm pert}^G\left((m_c^2/m_b^2);\,\mu\right)\cdot \frac{\mu_G^2}{m_b^2}\,+\,
...\right]
\label{np}
\end{equation}
where $\mu_\pi^2$ and $\mu_G^2$ are the expectation values of the kinetic and
chromomagnetic operators normalized at point $\mu$. The chromomagnetic operator
does not mix with the unit one due to the different spin structure, and
therefore its perturbative matrix element over heavy quark vanishes. As was
mentioned just above, in the
leading BLM approximation the perturbative value of $\mu_\pi^2$ vanishes as
well if one introduces the cutoff in the way adopted here. Thus one immediately
concludes that the contribution of the infrared region to the
perturbative coefficient of the leading operator in Eq.~(\ref{np}),
$a^{(0)}_{\rm pert}$, is suppressed by at least the third power of $\mu/m_Q$
to any order
in $\alpha_s$ in the BLM approximation~\footnote{This exactly corresponds to
the result of Ref.~\cite{bbz} which extended the statement of Ref.~\cite{pole}
about the absence of terms $\sim 1/m_Q$ through terms $\sim 1/m_Q^2$; however,
the latter seems to be a feature of the BLM approximation only.}. The
BLM-type calculations of $a^{(0)}_{\rm pert}(\mu)$ are then very infrared
stable.

We see that the contribution of the domain below $\mu$ to the
perturbative correction factor expressed in terms of $m_Q(\mu)$ is given
by the corresponding perturbative contribution to the matrix elements of
higher dimension operators only whose corrections to width scale
like $1/m_Q^3$ and
higher powers of the inverse mass. On the other hand, nonperturbative
corrections in the width are so far accounted for only through terms $\sim
1/m_Q^2$. Thus as long as at the scale $\mu\simeq 1\,\rm GeV$ the
perturbative part of matrix elements is dominated by the
condensate-like effects, the neglection of the remaining perturbative
corrections from the domain $k^2<\mu^2$ is not only legitimate, but rather
necessary.

Let us note that in the exclusive $B\rightarrow D^{(*)}$ transition the
dependence of the zero recoil amplitude on the masses enters only via the
perturbative (and also nonperturbative) corrections. Strictly speaking, even
here one should have used the running masses to calculate, say, the
perturbative factors $\eta_{A,V}$. Then, for example, to the second order in
$\alpha_s$ one would have
\begin{equation}
\eta_A \rightarrow \eta_A + c_m\frac{\alpha_s}{\pi}
\frac{\mu}{m_c}\,y(1-y)
\frac{d\eta_A}{dy} \simeq
\eta_A - c_m\left(\frac{\alpha_s}{\pi}\right)^2
\frac{\mu}{m_c}\left(1+y + \frac{2y}{1-y}\log{y}\right)
\label{eta}
\end{equation}
$$
y=m_c/m_b\;\;\;,\;\;\;\;\;\mu\ll m_c\;\;.
$$
The effect, however, does not appear in the BLM approximation;
numerically it is insignificant compared to nonperturbative
corrections~\footnote{From the theoretical perspective, however, this
modification is profound: it represents the only linear in $1/m_Q$ correction
to $\eta_A$ as it is defined in HQET. To phrase it differently: unless one
uses the running masses, i.e. introduces the corrections similar to those in
Eq.~(\ref{eta}), the known statements about the absence of corrections linear
in $1/m_Q$ to the zero recoil amplitudes from the low energy domain, are
incorrect. (For the most general justification of the above conclusion see
\cite{optical}, part 7.) The last term in Eq.~(\ref{eta}) thus represents the
counter-example to the Luke's theorem as it is formulated in HQET, see, e.g.
\cite{neubpr}; this effect appears only beyond the BLM approximation, at the
order $\alpha_s^2$ and has not been noticed so far.}.
The peculiarity of the determination of $|V_{qb}|$ from the total semileptonic
width is that the latter in the tree approximation is proportional to the
power of the quark masses, and therefore it does require the accurate
treatment of the heavy quark masses.

\section{Implications for Determination of $|V_{cb}|$ and $|V_{ub}|$}

Let us briefly discuss the phenomenological implications. Eliminating the
spurious $1/m_Q$ renormalon which, in fact, is not present in the relation
between
different observables in QCD \cite{pole} by using the running quark masses
instead of the pole masses, makes the
perturbative corrections essentially smaller.
Now we would get for $m_c/m_b=0.3$ and $m_b=4.8\,\rm GeV$
$$
\Gamma_{\rm sl}(b\rightarrow u) \simeq \frac{\tilde{m}_b^5
|V_{ub}|^2}{192\pi^3}
\left(1-0.62\frac{\alpha_s(m_b)}{\pi}\;-\;1.1
\left( \frac{\alpha_s}{\pi}\right)^2 \right)
$$
\begin{equation}
\Gamma_{\rm sl}(b\rightarrow c) \simeq \frac{\tilde{m}_b^5
|V_{cb}|^2}{192\pi^3}
\left(1-0.86\frac{\alpha_s(\sqrt{m_c m_b})}{\pi}\;+
\;1.7 \left( \frac{\alpha_s}{\pi}\right)^2
\right)
\label{21}
\end{equation}
$$
\tilde{m}_b=m_b(1.3\,{\rm GeV})\;\;,\;\;\;\; \tilde{m}_c=m_c(1.3\,\rm GeV)\;\;.
$$

Addressing the CKM mixing parameters $V_{ub}$ and $V_{cb}$, one needs to
consider
the square root of the widths, and in particular the corresponding perturbative
factors $\eta_\Gamma=\left(\Gamma_{\rm sl}^{\rm pert}/\Gamma_{\rm
sl}^{0}\right)^{1/2}$ which are directly related to extraction of $|V_{qb}|$.
In terms of $\tilde{m}$ they then look as follows:
$$
\tilde\eta_\Gamma\simeq
\left(1-0.31\frac{\alpha_s(m_b)}{\pi}\;-\;
0.6\left(\frac{\alpha_s}{\pi}\right)^2
\right)\;\;\;\mbox{ for }\;\;b\rightarrow u
$$
\begin{equation}
\tilde\eta_\Gamma\simeq
\left(1-0.43\frac{\alpha_s(\sqrt{m_c m_b})}{\pi}\;+\;
0.8\left(\frac{\alpha_s}{\pi}\right)^2
\right)\;\;\;\mbox{ for }\;\;b\rightarrow c
\label{22}
\end{equation}
These approximate expressions show the real sensitivity of extraction of
$|V_{ub}|$ and $|V_{cb}|$ to the
perturbative effects. For example, the impact of
the second order BLM correction is smallish. Moreover, because now the values
of
the heavy quark masses are essentially uncorrelated with $\alpha_s$, the first
order coefficient indicates the actual uncertainty in $|V_{cb}|$ associated
with the precise value of $\alpha_s$, which thus appears to be small and does
not exceed the one occurring
for the exclusive $B\rightarrow D^*$ zero recoil amplitude.\vspace*{.25cm}

Using the most recent evaluation of $m_b$ we can now obtain the updated
determination of $|V_{cb}|$ and the expression for $|V_{ub}|$; although they
are not really affected by the second order BLM-type corrections, we will use
the calculated terms \cite{wise} literally to obtain the `central' theoretical
values. Two clarifying comments are in order beforehand.

The above numerical values of the coefficients were shown for illustrative
purposes only; in fact, the ratio $m_c/m_b$ even for the one loop pole masses
appears to be less than $0.3$ when the $1/m_Q$ expansion relations (see
Eq.~(\ref{23}) below) are respected; this ratio decreases further for the
running masses. In our numerical analysis we compute the values of the
coefficients anew for proper values of $m_b$ and $m_c/m_b$.

Second comment concerns the value of $m_c$. In principle, it can be accurately
determined independently from charmonium sum rules, or from the semileptonic
$b\rightarrow c $ spectrum itself \cite{prl,Volmc}.
At present, however, the most accurate estimate follows from the relation
\begin{equation}
m_b-m_c=\frac{3M_{B^*}+M_{B}}{4}-\frac{3M_{D^*}+M_{D}}{4}+
\frac{\mu_\pi^2}{2}\left(\frac{1}{m_c}-\frac{1}{m_b}\right) +{\cal
O}\left(\frac{1}{m_Q^2}\right)
\label{23}
\end{equation}
which has been used for accurate determination of $|V_{cb}|$ in
Refs.~\cite{vcb,optical}. This relation is formally valid for masses
normalized at any legitimate point $\mu\gg \Lambda_{\rm QCD}$, and for our
purposes it would be most convenient to take $\mu$ directly the
same as used in Sect.~1, $\mu \sim 1\,\rm GeV$.
Strictly speaking, the result depends to some extent on the
particular choice of $\mu$ due to terms $\sim
(\alpha_s/\pi)\, \mu^3/m_Q^2$ (and higher in $\alpha_s/\pi$ and/or $\mu/m_Q$)
contributing to  Eq.~(\ref{23}) in the form of the perturbative pieces of
expectation values of higher dimension operators. In particular, one may be
concerned with the terms $\sim \alpha_s\,\mu^3/m_c^2$ for large values of $\mu
\simeq m_c$. In fact, this effect is easily controlled to all orders in the
framework of the BLM approximation,
$$
m_b(\mu')-m_c(\mu')=m_b(\mu)-m_c(\mu)+
\int_{\mu^2}^{\mu'^2}\, dk \frac{\alpha_s(k)}{\pi}
\left({\cal F}_m\left(\frac{k^2}{m_c^2}\right)-
{\cal F}_m\left(\frac{k^2}{m_b^2}\right)\right)\simeq
$$
$$
\simeq
m_b(\mu)-m_c(\mu)+\frac{\alpha_s(\lambda)}{\pi}\,\int_{\mu^2}^{\mu'^2}\,
dk\:\left({\cal F}_m\left(\frac{k^2}{m_c^2}\right)-{\cal F}_m
\left(\frac{k^2}{m_b^2}\right)\right)
\;+
$$
\begin{equation}
+ \;
\frac{b}{2}\left(\frac{\alpha_s(\lambda)}{\pi}\right)^2
\int_{\mu^2}^{\mu'^2}\, dk\:\log{\frac{\lambda}{k}}
\left({\cal F}_m\left(\frac{k^2}{m_c^2}\right)-
{\cal F}_m\left(\frac{k^2}{m_b^2}\right)\right)\;+\;...
\label{24}
\end{equation}
The variation of the mass difference does not exceed $20\,\rm MeV$
even when $\mu'$ varies up to $1.4\,\rm GeV$. This dependence on the
normalization point is less than the error
associated with the existing uncertainty in $\mu_\pi^2$ and can be
discarded \footnote{The suppression of corrections to the mass difference
reflects the fact that they start with terms $\sim (\mu/m_Q)^3$; this follows
from the absence of the perturbative contribution to $\mu_{\pi}^2$ in the BLM
approximation.}.
For the sake of definiteness we use the relation (\ref{23})
at the scale $0.5\,\rm GeV$; this literally decreases the value of
$m_b(\mu)-m_c(\mu)$ obtained without radiative corrections by only
a few $\rm MeV$.

For our numerical estimates we use, as the central value, the strong coupling
determined from the sum rules for $b\bar b$ production \cite{Volmb} as well:
\begin{equation}
\alpha_s^{\overline{\rm MS}}(1\,\rm GeV)=\alpha_s^V(2.3\,\rm GeV)=0.336\pm
0.011
\label{25}
\end{equation}
and the value \cite{Volmb}
\begin{equation}
m_b^* \simeq m_b(1.3\,\rm GeV) \simeq
m_b^{\rm pole}-0.56\alpha_s^V(2.3\,\rm GeV)\cdot
1\,\rm GeV=4.639\,\rm GeV\;\;;
\label{26}
\end{equation}
in view of the numerical close proximity of the SV scale $\sqrt{m_cm_b}$ to the
above $V$ scheme scale we
expressed the perturbative corrections in terms of $\alpha_s^V(2.3\,\rm GeV)$
in the case of $b\rightarrow c$ decays irrespective of the exact values of
the running quark masses, and for $b\rightarrow u$ decays use
$\alpha_s^V(4.8\,\rm GeV)$; the second order coefficients are then adjusted
appropriately. Again, this does not lead to any noticeable variation as
compared to different choices.

We then obtain for $\mu_\pi^2=0.5\,\rm GeV^2$ using the normalization scale
$\mu=1\,\rm GeV \,$:
$$
|V_{ub}|=0.00458\left(\frac{{\rm Br}(B\rightarrow X_u\ell\nu)}{0.002}
\right)^{\frac{1}{2}}\left(\frac{1.6\,\rm ps}{\tau_B}\right)^{\frac{1}{2}}
$$
and
\begin{equation}
|V_{cb}|=0.0408\left(\frac{{\rm Br}(B\rightarrow X_c\ell\nu)}{0.105}
\right)^{\frac{1}{2}}\left(\frac{1.6\,\rm ps}{\tau_B}\right)^{\frac{1}{2}}
\label{26a}
\end{equation}

It is important to emphasize that neither of the numbers above depend
essentially on the exact value of the scale $\mu$: varying it from $0$ (i.e.
using the ``two loop pole mass'') to $1.35\,\rm GeV$ literally changes the
value of $|V_{cb}|$ by only $-1.2$ percentage points. Using the running
masses is only instructive in showing explicitly the small effect of higher
order corrections; it is not mandatory in practice as long as one uses the
same approximations in the theoretical expression for the widths and for
determination of quark masses \cite{optical}. The dependence is more
pronounced in the case of $b\rightarrow u$, but even here $|V_{ub}|$
increases by only $2$ percentage points when $\mu$ is descended from $2\,\rm
GeV$ down to  $0.5\,\rm GeV$. Let us finally note that discarding the second
order perturbative terms $\sim b(\alpha_s/\pi)^2$ altogether would literally
lead to the explicit coefficients $0.00460$ and $0.0410$ in the above
equations, respectively, provided one uses the same running masses normalized
at $1\,\rm GeV$. The change is less than a half of percentage
point~\footnote{Comparing with the previous estimate \cite{vcb}, the central
value of $|V_{cb}|$ obtained there for the same input parameters
$\Gamma_{\rm sl}(b\rightarrow c)$ and $\mu_\pi^2$ as adopted in the present
paper, reads $|V_{cb}|=0.0405$. In this respect it is worth clarifying that
the numerical estimate in Ref.~\cite{vcb}, Eq.(25), was obtained using the
$V$ scheme coupling corresponding to the $\overline{MS}$ one quoted in
\cite{vcb}; the latter was $15\%$ larger than the central value adopted in the
present paper.}!

Turning to the estimate of the actual theoretical accuracy of extracting
$|V_{cb}|$ we note that the dominant uncertainty comes from the value of
$\mu_\pi^2$: varying it by $\pm 0.1\,\rm GeV^2$ changes $|V_{cb}|$ by the
factor
$(1\mp 0.013)$ which, in fact, comes mainly from the related variation of
$m_b-m_c$. The dependence on the exact value of $m_b^*$ is rather
weak: changing it by $\pm 30\,\rm MeV$ leads to the shift in $|V_{cb}|$ by
$\mp 0.6$ percentage points.

The main uncertainty in $|V_{cb}|$ is associated with the exact value of
$m_b-m_c$. Changing it by  $\pm 30\,\rm MeV$ as compared to the one given by
Eq.~(\ref{23}) leads to the factor $(1\mp 0.011)$. The relation (\ref{23})
is exact to
the order $1/m_Q^2$, however it may well be affected at the level of
$20\,\rm MeV$ by terms $\sim 1/m_c^3$. Therefore, at a percent level of
accuracy one needs to estimate $1/m_Q^3$ terms in the heavy quark expansion if
relies on this mass relation for fixing $m_c$. On the other hand, the
value of $m_b-m_c$ can be accurately determined from the semileptonic spectrum
itself not appealing to the expansion in $1/m_c$; then the nonperturbative
corrections calculated \cite{prl} through terms $1/m_b^2$ are sufficient
to determine the mass difference with the necessary accuracy \cite{Volmc}, and
even when $m_c$ is not large. Combining the two methods one can, therefore,
eliminate this source of uncertainty in $|V_{cb}|$
at a percent level in the both limits of
large and small $m_c$, i.e. have the determination of $|V_{cb}|$ limited,
on theoretical side, only by terms $\propto 1/m_b^3$.

Finally, the dependence on the exact value of $\alpha_s$ is also rather
moderate: changing $\alpha_s^{\overline{\rm MS}}(1\,\rm GeV)$ by $\pm 0.02$
emerges in the variation of $|V_{cb}|$ by $\pm 0.6$ percentage points only.

Even better stability would hold for $|V_{ub}|$: the dependence on
$\mu_\pi^2$ is very weak here, $\pm 0.12$ percentage points for every $\pm
0.1\,\rm GeV^2$ in the latter. The similar change in $m_b^*$ by $\pm 30\,\rm
MeV$ generates the variation by the factor $(1\mp 0.016)$ and the uncertainty
of $\pm 0.02$ in $\alpha_s^{\overline{\rm MS}}(1\,\rm GeV)$ translates into
the factor $(1\pm 0.009)$.

The fit error bars quoted in Ref.~\cite{Volmb} are
\begin{equation}
\delta m_b^*=\pm 0.002\,{\rm GeV}\;\;,\;\;\;\delta \alpha_s^{\overline{\rm
MS}}(1\,{\rm GeV})=\pm 0.011.
\label{27}
\end{equation}
If used literally they would lead only to the negligible error in $|V_{cb}|$
less than $0.3\%$.
It is clear, however, that for our purposes it cannot be
taken at
face value, even leaving aside the presently unknown value of $\mu_\pi^2$.
There are other sources of $(\alpha_s/\pi)^2$ perturbative corrections, both in
the inclusive width and in the determination of $m_b^*$ in Eq.~(\ref{26})
which are probably more important. Of course, the above quoted
error in $m_b$ is not relevant here
as well, because $m_b^*$ cannot be viewed as the exact value of the running
mass at a known scale.
To be on the conservative side we feel necessary, before the dedicated analysis
\cite{Volrun} for the running mass is completed,
assign much larger uncertainty to $m_b(1\,{\rm GeV})$, $\delta m_b\approx
50\,\rm MeV$ (the uncertainty in this running mass in the analysis of sum rules
for $b\bar b$ production seems to be about $20\div 30\,\rm MeV$
\cite{Volrun}),
and allow for the uncertainty in $a_2$ for the inclusive widths of at
least $\pm (3\div 5)$ as long as the complete two loop calculations for the
width
are not available.
Adding these theoretical error bars one ends up with the current theoretical
accuracy in $|V_{cb}|$ of about
$(3\div 3.5)\%$ as a rather conservative estimate,
provided $\mu_\pi^2$ is known. Anyway, it is quite possible that the
theoretical precision cannot be reliably
pushed below a percentage level due to potential preasymptotic
corrections given by ``exponential'' terms limiting the applicability of
duality for semileptonic decays of actual $b$ hadrons \cite{shif}.

The similar theoretical accuracy of the hypothetical determination of
$|V_{ub}|$ from the inclusive
width $\Gamma(B\rightarrow X_u \ell\nu)$ might seem
to be extraordinary good, better than $3\%$. In fact,
rather sizable effects
here may, in principle, come from terms $\sim 1/m_b^3$, in particular
due to generic Weak Annihilation processes \cite{WA} possible in the
KM suppressed semileptonic decays, which are expressed in terms of the
expectation
values of local four fermion operators \cite{VSold,BUold,WA,dike}; their effect
can be numerically enhanced here. To some extent it can be controlled
by studying semileptonic KM suppressed decays separately for charged and
neutral $B$
mesons, and in particular in the end point region where the effect mainly
originates from \cite{WA}. Therefore the
determination of
$|V_{ub}|$ based only on the high energy part of the lepton
spectrum can undergo
strong higher order nonperturbative corrections that are not well known yet.
If the total $b\rightarrow u $ semileptonic width were accurately known then
the uncertainty in $|V_{ub}|$ would not exceed $5\%$ level.

\section{Conclusions}

We have pointed out in this paper that the large (in
particular for the $b\rightarrow u$ channel) second order perturbative
coefficients for the inclusive semileptonic widths of beauty particles are
mainly associated with the similar contributions to the pole masses of heavy
quarks when the widths are expressed in terms of $m_Q^{\rm pole}$. The
corrections become
small if one uses theoretically well defined running masses normalized at the
scale about $1\,\rm GeV$; moreover, only these masses can and, as a matter of
fact, are determined from experiment with necessary accuracy. In particular,
the systematic shift in the values of $|V_{cb}|$ and $|V_{ub}|$ expressed in
terms of the corresponding semileptonic widths, constitutes less than one
percentage point when proceeding from the first order perturbative expressions
to the ones where the $\alpha_s^2$ corrections are calculated using the BLM
approximation, provided the value of $m_b(1\,\rm GeV)$ is fixed. Thus the
perturbative corrections, as well as nonperturbative effects, seem to be under
good control at the level corresponding to a percent relative
accuracy in $|V_{qb}|$.

Even in the semileptonic decays of charm
the consistent OPE treatment leads to the
smallness of perturbative corrections: the large negative value of perturbative
corrections found in Refs.~\cite{wise} in fact came from the momentum region
near and below
the infrared pole in the strong coupling. This contribution is therefore
numerically irrelevant and is to be excluded from the calculations applied to
charm; it is completely accounted for by nonperturbative
effects within Wilson OPE. For
this reason we cannot consider convincing the conjecture stated in
Refs.~\cite{wise} that the apparent numerical discrepancy of experimental
semileptonic width of $D$ mesons with theoretical expectations
is associated with the uncontrollable nature
of the perturbative series. The possibility to witness sizable violations of
duality in this case \cite{dike} seems to be more probable.

The theoretical accuracy of calculations of total semileptonic widths of $b$
particles appears to be very good when the input from the analysis of the
$b\bar b$ threshold domain \cite{Volmb} is used. One has, as the central
values, at $\mu_\pi^2=0.5\,\rm GeV^2\,$
$$
|V_{ub}|=0.00458\left(\frac{{\rm Br}(B\rightarrow X_u\ell\nu)}{0.002}
\right)^{\frac{1}{2}}\left(\frac{1.6\,\rm ps}{\tau_B}\right)^{\frac{1}{2}}\;\;,
$$
\begin{equation}
|V_{cb}|=0.0408\left(\frac{{\rm Br}(B\rightarrow X_c\ell\nu)}{0.105}
\right)^{\frac{1}{2}}\left(\frac{1.6\,\rm ps}{\tau_B}\right)^{\frac{1}{2}}\;\;.
\label{26b}
\end{equation}
The main uncertainty in $|V_{cb}|$ at present comes from the exact value of
$\mu_\pi^2$:
\begin{equation}
|V_{cb}|\propto \left(1-0.013\frac{(\mu_\pi^2-0.5\,\rm GeV^2)}{0.1\,\rm
GeV^2}\right)\cdot \left(1-0.006\frac{\delta m_b^*}{30\,\rm MeV}\right)\;\;.
\label{new}
\end{equation}
The conservative estimate of other uncertainties associated, in particular,
with not yet calculated part of the second order perturbative corrections,
is about $3$ percentage points. The above expressions for $|V_{cb}|$
practically do not differ from the previous estimate obtained in
Ref.~\cite{vcb}.

The theoretical accuracy of calculating $\Gamma_{\rm sl}(b\rightarrow u)$ is
even better; there is no strong dependence on $\mu_\pi^2$ here, and the
precision is better than $5\%$ of the equivalent relative variation in
$|V_{ub}|$. The uncertainty associated with the mass of $b$ quark does not
exceed $2\div 3$ percentage points and is not dominant. \vspace*{.2cm}

The analysis made in this paper suggests also
that the impact of the third and higher
order perturbative corrections, which can be computed straightforwardly within
the BLM approximation \cite{smvol,bb2,bbb},
is to be small and under good theoretical control provided one
treats the quark masses in the consistent way. Namely, if the pole mass for
$m_b$ is used, its numerical value must be calculated in exactly the same
approximation as applied to the calculation of the width. For example, one can
impose the constraint that the one (or two) loop pole mass is not changed by
the higher order corrections. More simple, and theoretically appropriate,
approach is to express the corrections in terms of the running mass at the
scale around $1\,\rm GeV$, which can be even better determined from
experiment.
(The usual technical problem with the precise definition of
normalization point is absent in the BLM approximation.)
The BLM corrections can then effectively sum the potentially
dominant terms coming from {\em high} momenta by taking into account running of
$\alpha_s$.
Because the coupling in this region is relatively
small, the running is not sharp and is well approximated already by the second
term in the expansion of $\alpha_s$. For this reason one expects numerically
small impact of the BLM corrections beyond $b(\alpha_s/\pi)^2$ terms, and
the
obtained series must be under good numerical control. On the contrary, the
effect of the low momentum physics is to be
computed by means of the Wilson OPE rather than
summing up perturbative series unstable in infrared.
\vspace*{0.3cm}

{\em Note added: } When this paper was in progress I was informed by V.~Braun
about computations of the effects of higher order perturbative
corrections for semileptonic widths made within the technique of
papers~\cite{bb2,bbb}, Ref.~\cite{bbb2}.
The preliminary results reported agreed with the
expectations that the actual impact of higher order corrections is smallish
when the proper corrections in the quark masses are introduced.

\vspace*{0.3cm}

{\bf ACKNOWLEDGMENTS:} \hspace{.4em} Numerous discussions of
problems touched upon in this paper, both of theoretical and practical nature,
with I.~Bigi, M.~Shifman,
A.~Vainshtein and M.~Voloshin are gratefully acknowledged. I am grateful
to V.~Braun for useful exchange of ideas and keeping me informed about the most
recent results.
This work was supported in part by DOE under the grant
number DE-FG02-94ER40823.

\vspace*{1cm}

\begin{center}
\begin{tabular}{|c|c|c|c|c|}\hline
  & $b\rightarrow c$ & $b\rightarrow c$ & $b\rightarrow u$ &
$b\rightarrow u$ \\ \hline
$\mu\,,\rm GeV$
& $\tilde{a}_1$ & $\tilde{a}_2$ & $\tilde{a}_1$ & $\tilde{a}_2$\\ \hline
$0.3$ & -1.50 & -1.9 & -2.00 & -12.6 \\
$0.5$ & -1.38 & -0.9 & -1.72 & -9.5  \\
$0.75$ & -1.23 & 0.1 & -1.38 & -6.3 \\
$1.0$ & -1.07 & 0.9 & -1.03 & -3.7  \\
$1.25$ & -0.89 & 1.6 & -0.70 & -1.5 \\
$1.5$ & -0.71 & 2.1 & -0.35 & 0.4 \\
$2.0$ & -0.32 & 2.8 & 0.32 & 3.5  \\ \hline
\end{tabular}
\end{center}
{\Large{Table 1:}} \hspace*{0.2em}
Dependence of the perturbative coefficients $\tilde{a}_1$
and $\tilde{a}_2$ on the scale $\mu$ for $m_b=4.8\,\rm GeV$ and
$m_c/m_b=0.3$. The strong coupling $\alpha_s$ is assumed to be
defined in the $V$ scheme and normalized
at $\,\sqrt{m_cm_b}\;$ for $b\rightarrow c$ and at $\,m_b\,$
for $\;b\rightarrow u\:$.

\end{document}